\documentstyle[preprint,aps,epsf]{revtex}
\newcommand{\beq}{\begin{equation}}
\newcommand{\eeq}{\end{equation}}
\newcommand{\ds}{\displaystyle}
\newcommand{\beqar}{\begin{eqnarray}}
\newcommand{\eeqar}{\end{eqnarray}}
\begin{document}
\draft
\title{ Statistical prefactor and nucleation rate near and
out of the critical point }
\author{Larissa V. Bravina$^{1,2,}$\footnote{Alexander von
Humboldt Foundation Fellow}~ and Eugene E. Zabrodin$^{1,2}$}
\address{
$^1$Institute for Theoretical Physics, University of Frankfurt,\\
Robert-Mayer-Str. 8-10, D-60054 Frankfurt, Germany \\
$^2$Institute for Nuclear Physics, Moscow State University,
119899 Moscow, Russia\\
}

\maketitle

\begin{abstract}
The nucleation rate derived in the classical theory contains at least
one undetermined parameter, which may be expressed in terms of the 
Langer first-principles theory. But the uncertainties in the 
accounting for fluctuation modes, which are either absorbed into the 
free energy of a critical cluster or not, result in different 
evaluations of the statistical prefactor and nucleation rate. We get 
the scaling approximations of the nucleation rate for the vapour 
condensation both near and out of the critical range. The results 
obtained deserve the experimental verification to resolve the 
theoretical uncertainty. 
\end{abstract}
%{\it Key words\/: Langer theory, fluctuation modes, nucleation rate,
%vapour condensation}

\widetext

The homogeneous nucleation occurred in metastable states has been area
of much theoretical and experimental activity since time of Maxwell 
and Van der Waals. The classical nucleation theory was worked out 
about 60 years ago mainly by Volmer, Becker and D{\"o}ring, Zeldovich,
and Frenkel \cite{Volm26,BeDo35,Zeld42,Fren46}. Under the treatment of 
relaxation of the initially homogeneous metastable state as a kinetic 
process, triggered by the creation of spherical clusters of the new, 
stable, phase as thermal fluctuations in the metastable phase, the 
classical theory provides us the fundamental nucleation rate 
\beq
\ds
I = I_0\, \exp\left( - \frac{\Delta F_c}{k_B T} \right)\ ,
\label{1}
\eeq
where $I_0$ is a pre-exponential factor, $\Delta F_c$ is the minimum
work needed to create critically large cluster of radius $R_c$, $k_B$
is Boltzmann constant, and $T$ is the temperature of the system. The
nucleation rate yields the number of viable clusters of the new phase 
per unit volume passing through the critical region per unit time via
the equilibrium number of critical clusters $f_0(R_c)$ given by the 
theory of thermal fluctuations. Usually \cite{Fren46,Kelt91} 
Eq.~(\ref{1}) is rewritten in the form
\beq
\ds
I = B(R_c)\, f_0(R_c)\, Z\ ,
\label{2}
\eeq
containing the size diffusion coefficient $B(R_c)$ and the Zeldovich
factor
\beq
\ds
Z = \left( \frac{\gamma}{2 \pi k_B T} \right)^{1/2}\ \ \ ,
\ \ \ \gamma = - \left( \frac{\partial^2 \Delta F(R)}{\partial R^2} 
\right)_{R = R_c} \ .
\label{3}
\eeq
This is the principal result of the classical theory. According to the 
thermodynamic theory of fluctuations the multiplicities of clusters of
various sizes obeys the Boltzmann (or rather Gibbs) distribution
\beq
\ds
f_0(R) = C_0\, \exp\left( - \frac{\Delta F(R)}{k_B T} \right) \ ,
\label{4}
\eeq
and the coefficient $C_0$ is not determined in the classical theory.
To find out the diffusion coefficient $B(R_c) \propto 1/(R - R_c) \,
({\rm d}R/{\rm d}t)_{R = R_c}$ one has to solve the macroscopic 
equations considering, for instance, the growth of a cluster due to 
the diffusion flux through its interface.

The coarse-grained field theoretical approach to the nucleation
phenomenon was developed by Langer \cite{Lang67,Lang69}, who
generalized the earlier works of Landauer and Swanson \cite{LaSw61},
and Cahn and Hilliard \cite{CaHi59}. The prefactor $I_0$ in the 
theory of Langer is shown to be a product of the dynamical prefactor 
$\kappa$ and the statistical prefactor $\Omega_0$
\beq
\ds
I_0 = \frac{\kappa}{2 \pi} \Omega_0 \ ,
\label{5}
\eeq
and both prefactors are determined explicitly in the theory. The 
dynamical prefactor is related to the growth rate of the critical
cluster, and the statistical prefactor is a measure of the phase
space volume available for the nucleation. Since the beginning of 
70's, the
Langer theory has been applied to describe first order phase 
transition in various systems including the vapor condensation
\cite{LaTu73,TuLa80}, nucleation in binary fluids \cite{Kawa75},
solidification of the melt \cite{GrGu85}, and hadronization of the 
quark$-$gluon plasma \cite{CsKa92,Zabr98}, produced in heavy$-$ion 
collisions.

As was shown by the authors recently \cite{BrZa97a}, nucleation rate 
(\ref{5}) of the field nucleation theory may be obtained under the
certain assumptions in a form identical to that of the classical 
theory. It means, particularly, that one is able to determine the
unknown constant $C_0$ in terms of the modern theory. However, in 
the present paper we show that the difficulties in the evaluation of 
$\Omega_0$ give rise to the uncertainty in the determination of the
nucleation rate, which can be clearly verified experimentally.

Our arguments are the following. According to the nucleation theory, 
both classical and modern, the relaxation of the homogeneous 
metastable state proceeds via the formation of clusters of the new,
stable, state. The free energy density functional $F(\{\eta\})$ of 
the system has two local minima in the configuration phase space
$\{\eta\}$, described by a set of $N$ collective coordinates 
$\eta_i$, corresponding to the states of metastable and stable 
equilibrium. These wells in the phase space are separated by the 
potential barrier related to the formation of cluster (or fluctuation)
of critical size. The nucleation rate is determined by the finite 
probability flux from the metastable configuration to the stable one 
across the vicinity of point with the lowest energy on the potential 
barrier, the so-called saddle point.

In the harmonic approximation for $F$ the statistical prefactor 
appeared in Langer theory reads \cite{GMS83} 
\beq
\ds
\Omega_0 =  V\, \left( \frac{2 \pi k_B T}{|\lambda_1|} \right)
^{1/2}\, \left[ \frac{{\rm det}(M^\prime / 2 \pi k_B T)}{{\rm det}
(M_0 /2 \pi k_B T)} \right]^{-1/2}\ .
\label{6}
\eeq
Here $V$ is the volume of the system, $M_0$ and $M$ denote the 
matrix $ M_{i j} = \partial^2 \Delta F / \partial \eta_i
\partial \eta_j$, evaluated at the metastable minimum $\{\eta_0\}$ 
and at the saddle point, and $\lambda_1$ is the only negative 
eigenvalue of $M$, associated with the instability of the critical 
cluster against expansion or shrinking. The prime in Eq.~(\ref{6}) 
indicates that the eigenvalue $\lambda_1$ is omitted. Then, the 
translational symmetry is broken because of the presence of critical 
cluster. These three translational modes with zero eigenvalues give 
rise to the factor \cite{Lang67} proportional to 
$ (|\lambda_1|)^{ -3/2}$, and Eq.~(\ref{6}) is transformed into
\beqar
\ds
\Omega_0 &=& {\cal V}\, \left( \frac{2 \pi k_B T}{|\lambda_1|} \right)
^{1/2}\, \prod_{\alpha=5}^N \left( \frac{2 \pi k_B T}{\lambda_\alpha^{
(S)}} \right)^{1/2} \, \prod_{\beta=1}^N \left( \frac{\lambda_\beta^{
(0)}}{2 \pi k_B T} \right)^{1/2} \ , \\ 
\label{7}
{\cal V} &=& V\, \left( \frac{8 \pi \sigma}{3 |\lambda_1|} 
\right)^{3/2}\ .
\label{8}
\eeqar
From here the uncertainties in the determination of the $\Omega_0$ 
begin. If the products over $\beta$ and $\alpha$ are absorbed 
\cite{Lang69} into the free energy of the metastable state and the 
saddle point, respectively, then Eq.~(\ref{7}) is reduced to
\beq
\ds
\Omega_0^{(1)} = {\cal V}\, \frac{32 \pi^2 (k_B T)^{1/2}}
{|\lambda_1|^2}\, \left( \frac{\sigma}{3} \right)^{3/2}\ ,
\label{9}
\eeq

If, on the other hand, one will pair \cite{LaTu73} the eigenvalues
$\lambda_{\alpha}^{(S)}$ and $\lambda_{\beta}^{(0)}$ so, that the 
corrections to the free energy difference remain of order $R_c^3 / V$ 
in the limit $V \rightarrow \infty$, there will be four unpaired 
eigenvalues $\lambda_{\beta}^{(0)}$'s at the bottom of the spectrum 
of eigenvalues
\beq
\ds
\lim_{R \rightarrow \infty} \prod_{\beta = 1}^4 \left( 
\lambda_\beta^{(0)} \right)^{1/2} = \left( \frac{1}{2 \pi k_B T}\,
\frac{\partial^2 f}{\partial n_g^2} \right)^2 \ ,
\label{10}
\eeq
where the second derivative of the Helmholtz free energy $f$ with 
respect to the local density $n_g$ is inversely proportional to the
correlation length $\xi$ squared, $\partial^2 f / \partial n_g^2 
\propto \xi^{-2}$. The statistical prefactor in the latter case turns
out to be
\beq
\ds
\Omega_0^{(2)} = {\cal V}\, \left( \frac{2}{3\sqrt{3}} \right)\,
\left( \frac{\sigma}{ k_B T} \right)^{3/2}\,
\left( \frac{R_c}{\xi} \right)^4\ .
\label{11}
\eeq

To make a comparison between the nucleation rates calculated with 
$\Omega_0^{(1)}$ and $\Omega_0^{(2)}$ one should know also the 
dynamical prefactor $\kappa$ and the minimum work needed to form the
critical cluster. For simplicity, let us consider the process of 
vapour condensation at the condensation point $T$, corresponding to
a certain pressure $p_0$. If the vapour is overcompressed with respect 
to $p_0$, the critical radius is determined by the expression 
\beq
\ds
R_c = \frac{2 \sigma}{\Delta \mu \Delta n} = 
\frac{2 \sigma v_l}{\Delta \mu} \ ,
\label{12}
\eeq
where $v_l$ is the volume per particle of the liquid phase, 
$\Delta n$ is the difference in local number densities, and 
$\Delta \mu$ is the difference in chemical potentials of the vapour
and liquid phase
\beq
\ds
\Delta \mu = \mu_v - \mu_l = k_B T\, \ln \frac{p}{p_0} \ .
\label{13}
\eeq
Ratio $p / p_0$ indicates the degree of supersaturation in the 
system. We see that in the capillary approximation the critical radius 
scales near the condensation point as $(\ln \epsilon)^{-1}$. Since for
the only negative eigenvalue $\lambda_1$ one gets $\lambda_1 = - 2 
\lambda_Z^2 k_B T$ \cite{BrZa97a}, 
where the similarity number \cite{BrZa95} is given by $\lambda_Z = 
(4 \pi \sigma / k_B T)^{1/2} R_c$, the statistical prefactor 
$\Omega_0^{(1)}$ should increase with the rise of $\epsilon$,
\beq
\ds
\Omega_0^{(1)} \propto (\ln \epsilon )^4\ ,
\label{14}
\eeq
while $\Omega_0^{(2)}$ drops inversely proportional to the same rate
\beq
\ds
\Omega_0^{(2)} \propto (\ln \epsilon )^{-4}\ .
\label{15}
\eeq

For the dynamical prefactor we have \cite{TuLa80,Kawa75}
\beq
\ds
\kappa = \frac{2 \lambda \sigma T v_l^2}{l^2 R_c^3} \propto
(\ln \epsilon)^3 \ .
\label{16}
\eeq
Here $\lambda$ is the heat conductivity and $l$ is the latent heat.
The excess of the free energy due to formation of the critical 
cluster is simply
\beq
\ds
\frac{\Delta F_c}{k_B T} = \frac{\lambda_Z^2}{3} = 
\frac{b_0}{(\ln \epsilon)^2}\ ,
\label{17}
\eeq
containing the parameter $b_0 = 16 \pi (\sigma / k_B T)^3 v_l^2 / 3$.

Combining our results we get that the nucleation rate is proportional
either to the factor $\alpha_1(\epsilon)$ or to the factor
$\alpha_2(\epsilon)$, correspondingly,
\beqar
\ds
I^{(1)}&=&\frac{\kappa}{2 \pi}\, \Omega_0^{(1)}\, \exp \left( -
\frac{\Delta F_c}{k_B T} \right) \propto \left( \ln \epsilon 
\right)^7\, \exp \left( - \frac{b_0}{(\ln \epsilon)^2} \right) 
\equiv \alpha_1(\epsilon)\ , \\
\label{18}
I^{(2)}&=&\frac{\kappa}{2 \pi}\, \Omega_0^{(2)}\, \exp \left( -
\frac{\Delta F_c}{k_B T} \right) \propto \left( \ln \epsilon 
\right)^{-1}\, \exp \left( - \frac{b_0}{(\ln \epsilon)^2} \right) 
\equiv \alpha_2(\epsilon)\ .
\label{19}
\eeqar

\begin{figure}[htp]
\centerline{\epsfysize=15cm \epsfbox{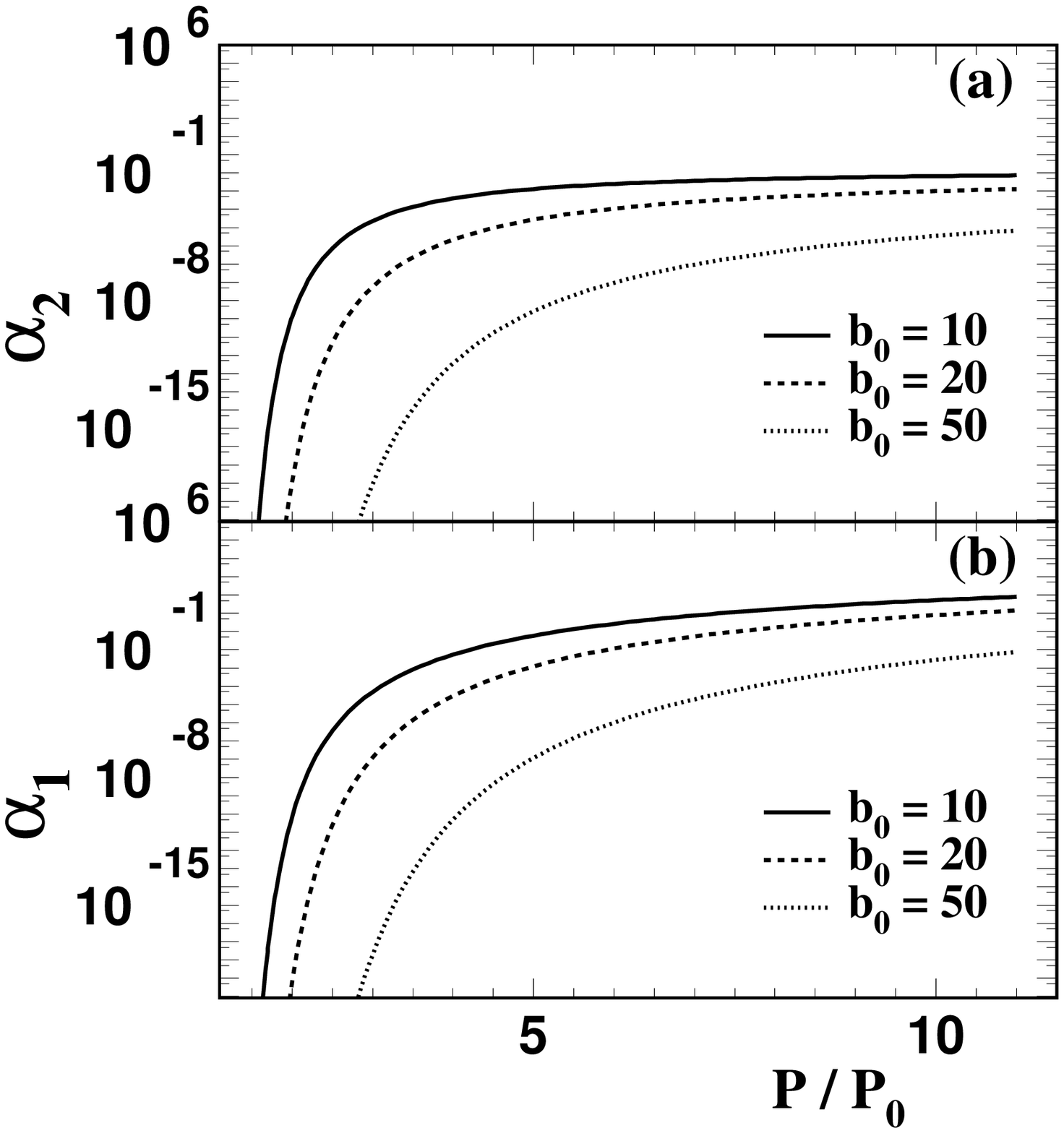}}
\caption{ 
Factors $\alpha_1$ and $\alpha_2$ as functions of the degree of
vapour supersaturation $p/p_0$ calculated with $b_0 = 10$ 
(solid lines), 20 (dashed lines), and 50 (dotted lines). }
\label{fig1}
\end{figure}

Figure~\ref{fig1} depicts the evolution of both $\alpha_1$ and 
$\alpha_2$ with the changing of the degree of supersaturation
$\epsilon$ for different values of the parameter $b_0$.
The rapid falloff of the exponential at $\epsilon \rightarrow 1$
dominates the increase of the statistical prefactor $\Omega_0^{(2)}$,
and although 
\beq
\ds
\lim_{\epsilon \rightarrow 1} \Omega_0^{(1)} = 0 \ \ \ {\rm and}\ 
\ \ \ \lim_{\epsilon \rightarrow 1} \Omega_0^{(2)} = \infty \ ,
\label{20}
\eeq
the nucleation rate in both cases drops to zero near the condensation 
point and rises up when the vapour is overcompressed. But the slopes
of ${\rm d} I / {\rm d} \epsilon$ are different, and this disagreement 
in theoretical predictions may be verified experimentally. Note that
in the region near the value $\epsilon = e = 2.7318 \ldots$ the 
slopes are similar. Therefore, the measurements should be carried out 
either with relatively weak supersaturations, or with the strong 
ones, say $\epsilon \approx 5$, which is still below the limit of 
supersaturation \cite{Zett69} obtained in the experiments on 
homogeneous nucleation in vapour.

Near the critical point we should take into account the power-law
approximations for the thermodynamic quantities appeared in the 
nucleation rates, and the critical indices come into play. These 
quantities are listed in Table~\ref{tab1} as functions of the 
standard dimensionless variable $\theta = 1 - T/T_c$ for the vapour 
side of the coexistence curve at $T < T_c$.  

From Table~\ref{tab1} it follows that the difference of the pressures 
$\Delta p$ inside and outside the droplet is 
$ \Delta p = \Delta n \Delta \mu \propto \theta^{\beta + \gamma}$ 
and, consequently, $R_c \propto \theta^{2 \nu - \beta - \gamma}$ 
and $\lambda_Z \propto \theta^{3 \nu - \beta - \gamma}$.

It is worth noting that $\lambda_Z \rightarrow 0$ as $\theta^{3/8}$
at $T \rightarrow T_c$. Therefore the formulae (\ref{9}) and 
(\ref{11}) cannot be applied to calculate the nucleation rate 
\cite{BrZa97b} because the criterion $k_B T / |\lambda_1| \ll 1$ of 
the steepest descent method is not fulfilled. To estimate the 
difference between the two approaches, however, one may apply the
droplet model of Fisher \cite{Fish67} instead of performing of 
numerical evaluation of the integrals. In the Fisher model the only
negative eigenvalue $\lambda_1$ is \cite{BrZa97b}
\beq
\ds
\lambda_1 = k_B T ( 9 \tau + 2 \lambda_Z^2)\ , 
\label{21}
\eeq
where $\tau \approx 2.2$ is the critical exponent. Although the 
critical radius $R_c^F$ determined in the Fisher model does not
coincide with that of the capillary approach, it has the same
scaling behaviour, $R_c^F \propto \theta^{2 \nu -\gamma -\beta}$,
in the proximity of the critical point. The free energy of a critical 
droplet is given by 
\beq
\ds
\frac{\Delta F_c^F}{k_B T} = \tau \left(1 - 3 \ln \frac{R_c^F}
{r_{min}} \right) - \frac{\lambda_Z^2}{3}\ ,
\label{22}
\eeq
where $r_{min}$ is the radius of a single molecule. This yields for
the exponential factor at $T \rightarrow T_c$
\beq
\ds
\exp \left( \frac{\Delta F_c^F}{k_B T} \right) \propto \theta^{
3 \tau ( \gamma + \beta -2 \nu )}\ .
\label{23}
\eeq
Substituting 
Eq.~(\ref{21}) in Eq.~(\ref{9}) we find that near the critical
point
\beq
\ds
\Omega_0^{(1)} \propto \theta^{3\nu}\ ,
%\Omega_0^{(1)} = {\cal V}\, \frac{32 \pi^2 (k_B T)^{1/2}}
%{(9 \tau)^2}\, \left( \frac{\sigma_0}{3} \right)^{3/2}\,
%\theta^{3\nu}\ ,
\label{24}
\eeq
and, consequently,
\beq
\ds
I^{(1)} = {\rm const}_1\, \times\, \theta^{2 \gamma + 3 \tau ( \gamma 
+ \beta -2 \nu )} \approx {\rm const}_1\, \times\, \theta^{3.6}\ .
\label{25}
\eeq

Similar estimations of the statistical prefactor $\Omega_0^{(2)}$
lead to the following parametrizations
\beq
\ds
\Omega_0^{(2)} \propto \theta^{15 \nu - 4 \gamma - 4 \beta}\ ,
\label{26}
\eeq
and
\beq
\ds
I^{(2)} = {\rm const}_2\, \times\, \theta^{12 \nu - 2 \gamma - 4 \beta
+ 3 \tau ( \gamma+ \beta -2 \nu )} \approx {\rm const}_2\, \times\, 
\theta^{5.1}\ .
\label{27}
\eeq
We see again that the resulting power-law approximation of the
nucleation rate $I^{(2)}$ as a function of $\theta$ deviates clearly
from that obtained for $I^{(1)}$.

In conclusion, we study the effect of absorption of the fluctuation
corrections to the excess free energy of a cluster into the cluster
activation energy. To compare different predictions made by the
Langer coarse-grained field theory of nucleation for the statistical
prefactor $\Omega_0$ we choose the liquid$-$gas phase transition near
and out of the critical point. It is shown that although the 
statistical prefactor drops to zero in the first case and goes to
infinity in the other one near the condensation point far from the
critical region, the qualitative behaviours of the nucleation rates 
are similar. However, the power-law parametrizations of the 
nucleation rate as a function of supersaturation are different. The
disagreement between the predictions is especially noticeable in the
region of very weak or, in contrast, strong supersaturations. 

The scaling behaviour of the statistical prefactor and total 
nucleation rate is studied also near the critical point in the Fisher 
droplet model. The qualitative dependences of $\Omega_0$ and $I$ on
the scaling parameter $\theta = 1 -T/T_c$ are similar in both 
approaches, while the analytic parametrizations are different again:
$\theta^{3.6}$ vs $\theta^{5.1}$. The experiments on homogeneous 
condensation of a vapour may certainly resolve this theoretical 
ambiguity.

{\bf Acknowledgments.}
We are indebted to the Institute for Theoretical Physics, University 
of Frankfurt for the warm and kind hospitality. One of us, L.B., 
acknowledges support of the Alexander von Humboldt Foundation.

\newpage

%\newpage
\mediumtext

\begin{table}
\caption{Power-law approximation of thermodynamic quantities near 
the critical point}

\begin{tabular}{ccl}
Quantity & Power-law approx.    & critical \\
         & at $T \rightarrow T_c$ & exponent \\
\tableline \tableline
$\sigma$   & $\sigma_0 \theta^{2 \nu}   $      & $\nu = 0.625$  \\
$\Delta n$ & $(\Delta n_0) \theta^\beta $      & $\beta = 0.5$  \\
$\lambda$  & $\lambda_0 \theta^{\nu - \gamma}$ & $\gamma = 1.0$ \\
$l    $    & $ l_0 \theta^\beta $      &   \\
$\xi  $    & $ \xi_0 \theta^{-\nu} $   &   \\
$\Delta\mu$& $ \propto \theta^\gamma $ &   \\
%$\Delta p $& $ \propto \theta^{\beta + \gamma}$&   \\
\end{tabular}
\label{tab1}
\end{table}

\end{document}